\documentclass[11pt]{article}

\usepackage{amsmath}
\usepackage{amssymb}
\usepackage{latexsym}
\usepackage{graphics}
\usepackage{psfrag,fancyhdr,epsfig}

\newcommand{\Xu}{\mathcal{X}}
\newcommand{\Yu}{\mathcal{Y}}

\newcommand{\Hu}{\mathcal{H}}

\newcommand{\Su}{\mathcal{S}}
\newcommand{\Tu}{\mathcal{T}}
\newcommand{\Qu}{\mathcal{Q}}
\newcommand{\Wu}{\mathcal{W}}
\usepackage{color}

\definecolor{orange}{rgb}{0.9,0.2,0}
\definecolor{brown}{rgb}{0.7,0.3,0.2}
\definecolor{fuxia}{rgb}{1,0,1}
\definecolor{skyblue}{rgb}{0,0.1,0.9}
\definecolor{violetred}{rgb}{0.8,0.13,0.56}
\definecolor{deeppink}{rgb}{1.00,0.08,0.5}
\definecolor{pink}{rgb}{1.00,0.75,0.80}
\definecolor{orchid}{rgb}{0.85,0.44,0.84}
\definecolor{lightpink}{rgb}{1.00,0.71,0.76}
\definecolor{bluish}{rgb}{0,0.6,0.8}
\begin{document}

\begin{titlepage}

\begin{centering}

\hfill hep-th/yymmnnn\\

\vspace{1 in}
{\bf {HIERARCHICAL NEUTRINO MASSES AND MIXING IN NON MINIMAL-$\bf{SU(5)}$} }\\
\vspace{1 cm}
{M. Paraskevas
 and K. Tamvakis}\\
\vskip 0.5cm
{$ $\it{Physics Department, University of Ioannina\\
GR451 10 Ioannina, Greece}}

\vskip 0.5cm

\vspace{1.5cm}
{\bf Abstract}\\
\end{centering}
\vspace{.1in}

 We consider the problem of neutrino masses and mixing within the
framework of a non-minimal supersymmetric SU(5) model extended by
adding a set of
${\underline{\bf{1}}},\,{\underline{\bf{24}}}$
chiral superfields accommodating three right-handed neutrinos. A
Type I+III see-saw mechanism can then be realized giving rise to a
hierarchical mass spectrum for the light neutrinos of the form
$m_3> m_2>> m_1$ consistent with present data. The extra colored
states are pushed to the unification scale by proton stability
constraints, while the intermediate see-saw energy scale and the
unification scale are maintained in phenomenologically acceptable
ranges. We also examine the issue of the large neutrino mixing
hierarchy $\theta_{23}\,>\,\theta_{12}\,>>\,\theta_{13}$ in the
above framework of hierarchical neutrino masses.

 \vfill

\vspace{2cm}
\begin{flushleft}

April 2011
\end{flushleft}
\hrule width 6.7cm \vskip.1mm{\small \small}
 \end{titlepage}

\section{Introduction}
The discovery of neutrino oscillations is the first encounter with
physics beyond the {\textit{Standard Model}}(SM). Data coming from various sources\cite{EXP} are
 conclusive that neutrinos are massive (at least two of
them) and that they mix, exhibiting oscillation
phenomena\cite{REV}. This implies a mismatch between flavour and
mass eigenstates in an obvious analogy with the CKM matrix in the
quark sector of the SM. In order to explain the new evidence on
the overall scale and structure of the neutrino mass matrix,
several proposals have been put forward, among which the most
interesting appears to be the so called {\textit{see-saw
mechanism}}\cite{SS}. It provides an elegant answer to the
smallness of the observed neutrino masses, although it leaves open
the issue of the underlying  structure of the neutrino mass
matrix. The see-saw mechanism is a general term and can be
realized in a number of forms and variations, with the basic idea
relying on the fact that a large energy scale $M>>m_W$ is
introduced through the coupling of the left handed neutrinos to a
sector of heavy fields. By integrating these heavy degrees of
freedom out an effective operator is produced giving small masses
to the neutrinos of order $\sim m_W^2/M$. A heavy scale in the
 neighborhood
 of $M\,\sim\,10^{14}\,GeV$ leads to an overall neutrino mass scale
 of $\,\sim\,10^{-1}\,eV$, in general agreement with observations. The usual classification of see-saw types in the literature is
based on the gauge properties of the heavy particles used to
mediate the see-saw mechanism\footnote{An alternative
classification based on the explicit mathematical expression for
the  light neutrino masses is also common.}. Types I, II, III
correspond to fermion singlets, scalar charged isotriplets and
fermion neutral isotriplets, respectively.

Since the see-saw mechanism requires a sector
of particles with masses well above the scales at which the validity
 of the SM has been established, it is natural to consider its realization
  in the framework of general approaches for the extension of the SM such
  as {\textit{Supersymmetry}} and {\textit{Grand Unification (GUTs)}}. The simplest choice of considering {\textit{minimal supersymmetric
$SU(5)$}} and introducing the required heavy fields as singlets,
 retains all the arbitrariness of realizing the see-saw mechanism
 within the SM without introducing any new constraint on scales
and structure. In addition, a phenomenologically viable scenario
within this context has to overcome the problems of the original
model such as proton decay and unrealistic fermion masses. A way around the
former would be to tune the Yukawa couplings and bi-unitary
transformations with the soft sector through certain relations, something unnecessary if the unification scale could be shifted. For realistic fermion masses the presence of nonrenormalizable operators would be required.

 In
order to obtain potentially interesting constraints on the scale
and structure of neutrino masses, the sector of heavy fields has
to partake in the GUT. This can be realized in other GUTs\cite{Buchmuller:2002xm}, such as
$SO(10)$ and
{\textit{flipped-$SU(5)$}}{\cite{RT}}, or by extending the gauge
non-singlet field content of
$SU(5)$. The realization of the so called {{type}}-$\bf{I}$ see-saw
mechanism in the SM introduces right-handed neutrinos as gauge
singlet fields. In contrast, in the {{type}}-$\bf{III}$
right-handed neutrinos are non trivially introduced as the neutral
components of isotriplet fields\cite{Foot:1988aq}. This can be
promoted to
 extended versions of $SU(5)$ that feature additional
chiral superfields in the ${\underline{\bf{24}}}$ representation,
each containing two suitable right-handed neutrino
candidates\footnote{ Fermions in a single $\underline{\bf{24}}$
representation have been introduced in the framework of non-supersymmetric $SU(5)$ in {\cite{Bajc:2006ia}}, where the see-saw mechanism  was realized with two right-handed neutrinos at a predicted low energy scale.}. A mixed ``type-$\bf{I+III}$"
see-saw mechanism can then be realized with an extra
$\underline{\bf{24}}${\cite{Biggio:2010me}}, while the most
appealing three generation scenario with three right-handed
neutrinos requires additional $\underline{\bf{24}}$'s or
$\underline{\bf{1}}$'s. In the present article we consider a
version of supersymmetric $SU(5)$ extended through the
introduction of extra chiral superfields
${\cal{S}}({\bf{1}}),\,{\cal{T}}({\bf{24}}),\,{\cal{T}}'({\bf{24}})$,
which provide us with three right-handed neutrino candidates. Our
basic assumption is that these right-handed neutrino fields obtain
a Majorana mass at a high but still intermediate scale a few
orders of magnitude below the
 unification scale. This assumption is supported
by a renormalization group analysis, incorporating proton lifetime
constraints{\cite{Murayama:2001ur}}, and allows for an
intermediate scale in the vicinity of $10^{13}\,-\,10^{14}\,GeV$.
Not all of the scales involved in the right-handed neutrino
Majorana mass matrix are constrained by the renormalization group.
Depending on assumptions, several possibilities emerge leading to
a different dependence of the resulting light neutrino masses on
these scales. Furthermore, the fact that two of the right-handed
neutrinos are members of the same $SU(5)$ representation leads to
a particular {\textit{rank 2}} structure of the resulting light
neutrino mass matrix that is accompanied by a massless eigenvalue.
Although this fact is modified by non-renormalizable terms, there
is a definite prediction for one superlight neutrino, not in
conflict with observations. Next, we examine the possibility of a
hierarchical light neutrino mass spectrum
$m_{\nu}^{(3)}\,>\,m_{\nu}^{(2)}\,>>\,m_{\nu}^{(1)}$. This can be
achieved in a variety of ways depending on assumptions either for
the mass scales involved or for the hierarchy of the Yukawa-type
couplings. We also consider whether the observed large neutrino
mixing can be accommodated in the framework of the
model{\cite{Barr:2000ka}}. We conclude that hierarchical
mixing patterns with
$\theta_{13}\,<<\,\theta_{12}\,\sim\,\theta_{23}$ can be obtained
with generic choices of Yukawa couplings exhibiting certain
structure.

\section{The Model}
The renormalizable part of the minimal $SU(5)$ superpotential, in terms of the chiral superfields  $\Qu ^c_i (10), \Qu _i(\overline{5}),
\Hu(\overline{5}), \Hu^c(5), \Sigma(24)$, is
\begin{eqnarray}\Wu_0 \,=\,
\Yu^{u}_{ij} \Qu^c_i \Qu^c_j\Hu^c +\Yu^{d} _{ij} \Qu_i \Qu^c _j
\Hu + \frac{M}{2} Tr(\Sigma^2)+
\frac{\lambda}{3!}Tr(\Sigma^3)+\lambda '\Hu^c \Sigma \Hu+M' \Hu^c
\Hu\,,
\end{eqnarray}
where we have suppressed $SU(5)$-indices and display only the
family indices $i,\,j$. Let us now introduce extra matter
supermultiplets $\Su(1),\Tu(24),\,\Tu'(24)$ with the standard
matter parity assignment\footnote{We have
${\cal{Q}},\,{\cal{Q}}^c,\,{\cal{S}},\,{\cal{T}}\,\rightarrow\,-1\,,$
while $\Sigma,\,{\cal{H}},\,{\cal{H}}^c\,\rightarrow\,1\,$.}. An
extra ${\cal{Z}}_2$ discrete symmetry, under which only $\Tu'(24)$
changes sign differentiates between them so that $\Tu'$ does not
couple to standard matter fields. The renormalizable contributions
of the new fields to the superpotential are
$$\Wu_1 \,=\, \Yu^{\Su}_i \Qu_i
\Hu^c \Su + \Yu^{\Tu}_i \Qu_i \Hu^c \Tu + \frac{\mu}{2} \Su^2  +
\,\frac{\mu '}{2} Tr(\Tu^2)\,+\, \frac{\mu''}{2}Tr({\Tu'}^2)\,$$
\begin{equation}\,+\,f\,Tr(\Tu^2\Sigma) + f'\,Tr(\Sigma
\Tu)\Su\,+\,f''Tr({\Tu'}^2\Sigma)\,\,.
\end{equation}
 The decomposition of the new matter multiplet $\Tu (24)$
 is
$$\Tu (24)=B(1,1,0)+T(1,3,0)+
O(8,1,0)+\Xu (3,2,-5/6)+\Xu^c(\overline{3},2,5/6)\,,$$
 where the $SU(3)\times SU(2) \times U(1)$ identification of each component is self-explanatory. Analogous is the decomposition of the primed field $\Tu'(24)$. Denoting by $T^0$ the neutral component of the isotriplet $T(1,3,0)$, we can identify the three right-handed neutrino candidates as $N^c_i=(\Su,B,T^0) $.

Symmetry breaking of SU(5) down to $SU(3)\times SU(2) \times U(1)$
is realized in the standard fashion through a non-zero vev of
$\Sigma$ in the direction $<\Sigma>=\frac{V}{\sqrt{30}} diag
(2,2,2,-3,-3)$. Note that the absence of cubic terms for the new
fields, due to their parity assignment, does not allow them to
acquire a non-zero vev and, thus, symmetry breaking proceeds
exactly as in the minimal case. All components of $\Sigma$ are
either {\textit{higgsed away}} or obtain masses of the order of
the GUT scale. The splitting between the masses of the Higgs
isodoublets $H_d,\,H_u$ and the Higgs coloured triplets $D,\,D^c$
contained in ${\cal{H}}\,=\,\left(\,H_d,\,D^c\right)$ and
${\cal{H}}^c\,=\,\left(\,H_u,\,D\right)$ is produced by the usual
fine-tuning $M'=\frac{3\lambda ' V}{\sqrt{30}}$, resulting in
massless doublets and superheavy triplets. Then, the effective
superpotential relevant for masses below the unification scale
$M_G$ reads
$$
\Wu_{eff}\,=\,Y^u_{ij} u^c_i Q_j H_u+ Y^d_{ij} d^c_i Q_j H_d+Y^e_{ij}
e_i^c L_j  H_d+ Y_i^{\Su} L_i \Su H_u+ Y_i ^B L_i B H_u+Y_i^{T}
L_i T H_u$$
$$+Y_i^{\Xu} d^c_i \Xu
H_u+\frac{M_{\Su}}{2}\Su^2+\frac{M_{B}}{2}B^2+M_{\Su B}\Su B
+\frac{M_{T}}{2}Tr(T^2)+M_{\Xu}\Xu\Xu^c+\frac{M_{O}}{2} Tr( O^2)$$
\begin{equation}\,+\,\frac{M_{T'}}{2}Tr({T'}^2)\,+\,\frac{M_{{O}'}}{2}Tr({O'}^2)\,+\,M_{\chi '}{\cal{X}}'{{\cal{X}}'}^c\,+\,\frac{M_{B'}}{2}{B'}^2
\,.{\label{W2}}\end{equation}
Matching the effective and the \textit{$SU(5)$-symmetric}
theory at $M_G$ leads to the following relations for the Yukawa
couplings
\begin{equation}
Y^u=2\Yu^u,\,\,\,\,\,\,\,\,\,\,\left(Y^e\right)^{\bot}=Y^d=\Yu^d,\,\,\,\,\,\,\,\,\,\,\,\,\,Y^{\Su}=\Yu^{\Su},\,\,\,\,\,\,\,\,\,\,Y^{\Xu}=Y^T=\frac{\sqrt{30}}{3}Y^{B}=\Yu^{\Tu}\,,\,
{\label{Y}}
\end{equation}
while for the mass parameters we get
\begin{equation}
M_\Su=\mu, \,\,\,\,\,\,M_B=\mu '-\frac{2fV}{\sqrt{30}}, \,\,\,\,\,\,\,\,\,\,
M_T=\mu '-\frac{6fV}{\sqrt{30}}\,,{\label{M1}}
\end{equation}
\begin{equation}
 M_O=\mu
'+\frac{4fV}{\sqrt{30}}, \,\,\,\,\,\,\,\,\,M_\Xu=\mu '-\frac{fV}{\sqrt{30}},\,\,\,\,\,\,\,\,\,\,\,\,\,M_{\Su B}=-f'V \,{\label{M2}}
\end{equation}
and
\begin{equation}
M_{T'}\,=\,\mu''\,-\frac{6f''V}{\sqrt{30}},\,M_{B'}\,=\,\mu''\,-\frac{2f''V}{\sqrt{30}},\,M_{O'}\,=\,\mu''\,+\,\frac{4f''V}{\sqrt{30}},\,M_{\cal{X}'}\,=\,\mu''\,-\frac{f''V}{\sqrt{30}}\,.{\label{M3}}
\end{equation}
 The see-saw scale is the scale of the right-handed neutrino mass matrix expressed in terms of the parameters
$M_{\Su},\,M_B,\,M_T$ and $M_{\Su B}$, related through the four
parameters $\mu,\,\mu',\,fV$ and $f'/f$. The allowed range for
these parameters will be strongly constrained by the requirements
of unification at a sufficiently high
scale. This will follow shortly from a renormalization group
analysis.

In addition to the renormalizable contributions above, non-renormalizable contributions to the superpotential
$${\cal{W}}_{NR}\,=\,\frac{\lambda_{IJKL}}{M_P}\,\Phi_I\Phi_J\Phi_K\Phi_L\,+\,O(1/M_P^2)\,+\,\dots.$$
can, in principle, affect masses, especially whenever we have
mass-degeneracies. We have denoted the scale  of
non-renormalizable interactions generically by $M_P$, expecting
their scale to be the Planck scale. The lowest order terms in
${\cal{W}}_{NR}$ are
$$
\Qu \Tu \Sigma \Hu^c + \Qu\Sigma \Hu^c \Su + \Tu \Qu^{c} \Hu \Hu +
\Qu^c \Qu^c \Sigma \Hu^c + \Sigma \Qu^c \Hu \Qu+ \Hu^c \Qu \Qu
\Hu^c +\Tu \Qu^c \Qu \Qu+\Qu^c\Qu\Qu \Su+$$
$$\Qu^c \Qu^c \Qu^c \Qu+\Tu^2 \Sigma^2
+\Sigma^2 \Tu \Su + \Hu \Tu^2 \Hu^c +\Sigma^2 \Su^2
+\Hu \Tu \Hu^c \Su + \Hu \Hu^c \Su^2+\Tu^4+ \Tu^3
\Su+ \Tu^2 \Su^2 + \Su^4  $$
\begin{equation}\,+\,\Sigma^4 + \Hu
\Sigma^2 \Hu^c+ \Hu \Hu^c \Hu \Hu^c\,+\,{\Tu'}^2\Sigma^2\,+\,\Hu{\Tu'}^2\Hu^c\,+\,{\Tu'}^4\,+\,{\Tu'}^2{\Tu}^2\,+\,{\Tu}{\Tu'}^2{\Su}\,+\,{\Tu'}^2{\Su}^2\,,{\label{NRW}}
\end{equation}
suppressing the factor $1/M_P$ and the dimensionless couplings  in
front of each term, all assumed to be of the same order. Among
these terms, those relevant for neutrino masses are the terms
${\cal{H}}^c{\cal{Q}}{\cal{Q}}{\cal{H}}^c$, leading to (tiny)
Majorana masses for left-handed neutrinos, the terms
${\cal{Q}}{\cal{T}}\Sigma{\cal{H}}^c$,
${\cal{Q}}\Sigma{\cal{H}}^c{\cal{S}}$, contributing to Dirac
masses, and the terms ${\cal{T}}^2\Sigma^2$,
$\Sigma^2{\cal{T}}{\cal{S}}$, $\Sigma^2{\cal{S}}^2$, contributing
to Majorana masses for the right-handed neutrinos.

\section{Energy Scales}
The sector of additional superfields ${\cal{T}},\,{\cal{T}}',\,{\cal{S}}$ carries with it a set of extra parameters, namely the mass parameters $\mu,\,\mu',\,\mu''$ and the couplings $f,\,f',\,f''$. A basic assumption of the model is that the Majorana mass of right handed neutrinos is at a high but still intermediate scale, a few orders of magnitude below $M_G$. Thus, we shall assume that the isotriplet component of ${\cal{T}}$ remains lighter than $M_G$. In addition, proton lifetime constraints translated to a high enough $M_G$ require the presence of an additional light color octet. These requirements correspond to new fine tunings of parameters, presumably, not worse than the standard GUT fine tunings. As a working set of choices, we take ($M_G^2=\frac{5g^2}{12}V^2$)
$$\mu'\,=\,(3\,-\epsilon)M_G/2,\,\,\,\mu''\,=\,(2+3\epsilon')M_G/5,\,\,\,\,\,f\,=\,\frac{5g}{4\sqrt{2}}\,(1-\epsilon),\,\,\,\,f''\,=\,-\frac{g}{2\sqrt{2}}\,(1\,-\epsilon')\,,$$
where $\epsilon\,\sim\,\epsilon'\,<<\,1$. These choices result in
\begin{equation}M_T\,=\,\epsilon\,M_G,\,\,\,M_{O'}\,=\,\epsilon'\,M_G\,,\end{equation}
while the rest of the masses are $M_O,\,M_{\cal{X}},\,M_{{\cal{X}}'},\,M_{T'}\,\sim\,O(M_G)$.

Thus, we assume that, apart from the MSSM fields and the color
octet and isotriplet superfields that have intermediate masses
$M_{O'}$ and $M_T$, all extra superfields decouple at $M_G$. In
addition, we assume that supersymmetry is broken at an
approximately common energy scale of ${m_S}\,=\,O(1\,TeV)$ at
which all superpartners decouple. From the one-loop
renormalization group equations for the three $SU(3)\times
SU(2)\times U(1)$ gauge couplings\footnote{The triplet-octet splitting has been previously studied for
SU(5) models at one and
two loops in {\cite{Dorsner:2006ye}} }, with the intermediate octet and
isotriplet mass scales inserted, we obtain the following
expressions for these couplings at $M_Z$
$$\frac{2\pi}{\alpha_3(M_Z)}\,=\,\frac{2\pi}{\alpha_G}\,-3\ln\left(\frac{M_G}{M_Z}\right)\,-4\ln\left(\frac{{m_S}}{M_Z}\right)\,+\,3\,\ln\left(\frac{M_G}{M_{O'}}\right)$$
$$\frac{2\pi}{\alpha_2(M_Z)}\,=\,\frac{2\pi}{\alpha_G}\,+\,\ln\left(\frac{M_G}{M_Z}\right)\,-\frac{25}{6}\ln\left(\frac{{m_S}}{M_Z}\right)\,+\,2\ln\left(\frac{M_G}{M_T}\right)$$
\begin{equation}\frac
{2\pi}{\alpha_1(M_Z)}\,=\,\frac{2\pi}{\alpha_G}\,+\,\frac{33}{5}\ln\left(\frac{M_G}{M_Z}\right)\,-\frac{5}{2}\ln\left(\frac{{m_S}}{M_Z}\right)\,,\end{equation}
where $\alpha_G$ is the common value of the three couplings at the
unification scale $M_G$. Inserting the existing recent
data{\cite{Nakamura:2010zzi}} for
$\alpha_3(M_Z),\,\alpha_2(M_Z),\,\alpha_1(M_Z)$, we obtain $M_G$
and $\alpha_G$, as well as the octet mass $M_{O'}$ for various
choices of the isotriplet mass treated as input. An octet mass
below $M_G$ sets a lower bound of $1.5\times 10^{16}\,GeV$ for the
unification scale. In {\textit{Figure 1}} we show the values of
$M_G$ obtained in terms of $M_T$. These values are tabulated in
{\textit{Table 1}} together with the corresponding values of
$M_{O'}$ and $\alpha_G$. Note that the values of $M_{O'}$ follow
$M_T$ within a close range, indicating an approximately common
intermediate scale. The values for $M_T$ in the proximity of
$10^{14}\,GeV$, corresponding to a safe $M_G\,\sim\,10^{17}\,GeV$,
have the correct order of magnitude required for the seesaw scale,
since $(10^2)^2/10^{14}\,\sim\,0.1\,eV$.
\begin{figure}[h!]
{\centerline{\includegraphics[width=8cm]{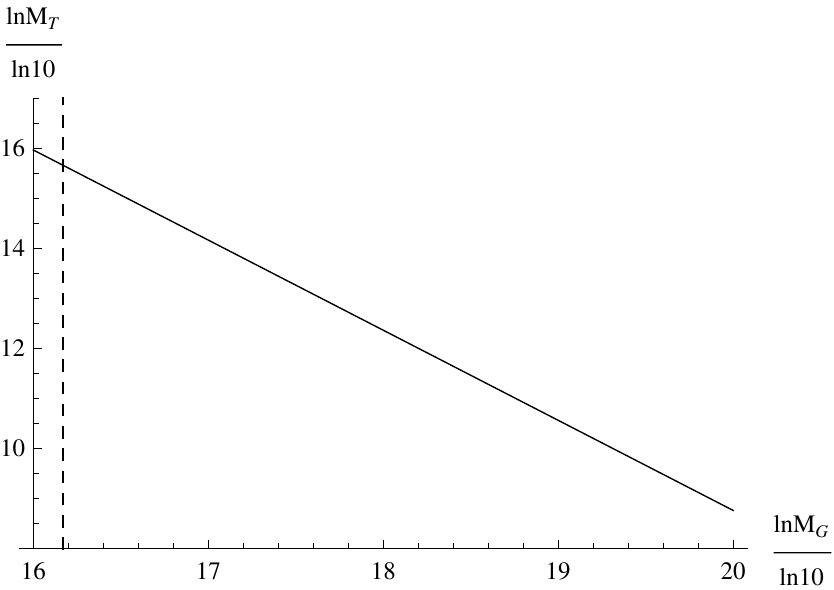}}}
\caption{ \small{Isotriplet mass  $M_T$ vs the unification scale $M_G$. The octet mass satisfying $M_{O'}\leqslant M_G$ sets a lower bound for unification at $M_G\approx 1.5\times 10^{16}\,GeV$.}}
\end{figure}

\begin{table}[h!]
{\centerline{\begin{tabular}{c c c c}
  \hline\hline
  $M_G$ & $M_{O'}$ & $M_T$ & $\alpha_G$  \\
  \hline
  $3\times 10^{16}$ & $3.1\times 10^{15}$  & $1.3\times 10^{15}$ & $0.04023$ \\
  $5\times 10^{16}$ & $1.0\times 10^{15}$  & $5.2\times 10^{14}$ & $0.04112$ \\
  $8\times 10^{16}$ & $3.6\times 10^{14}$  & $2.3\times 10^{14}$ & $0.04197$ \\
  $1\times 10^{17}$ & $2.2\times 10^{14}$ & $1.5\times 10^{14}$ & $0.04239$ \\
  $3\times 10^{17}$ & $2.0\times 10^{13}$ & $2.1\times 10^{13}$ & $0.04457$ \\
  $5\times 10^{17}$ & $6.4\times 10^{12}$  & $8.3\times 10^{12}$ & $0.04566$ \\
  $8\times 10^{17}$ & $2.3\times 10^{12}$  & $3.6\times 10^{12}$ & $0.04671$ \\
  $1\times 10^{18}$ & $1.4\times 10^{12}$ & $2.4\times 10^{12}$ & $0.04723$\\
  $3\times 10^{18}$ & $1.2\times 10^{11}$ & $3.3\times 10^{11}$ & $0.04996$\\
  \hline\hline
\end{tabular}}}
\caption{\small{Values ($GeV$) for the unification scale  $M_G$, the colored octet mass $M_{O'}$ and the weak isotriplet mass $M_T$. The corresponding unified coupling $a_G$ remains within the  perturbative limit.}}
\end{table}
$$\,$$

\section{Neutrino Masses}

The terms relevant for neutrino
masses can be easily singled out from the renormalizable part of
the superpotential ({\ref{W2}}).\footnote{{For simplicity, in our treatment of masses and
mixings we neglect CP-violation}} These terms are
\begin{eqnarray}
Y_i^{\Su} L_i \Su H_u+ Y_i ^B L_i B H_u+Y_i^{T} L_i T
H_u+\frac{M_{\Su}}{2}\Su^2+\frac{M_{B}}{2}B^2+M_{\Su B}\Su B
+\frac{M_{T}}{2}T^2
\end{eqnarray}
or
$$v_u\left(\,Y_i^{{\cal{S}}}\,{\cal{S}}\,+\,Y_i^BB\,-\frac{Y_i^T}{\sqrt{2}}\,\tau_0\,\right)\,\nu_i\,+\,\frac{M_{\cal{S}}}{2}{\cal{S}}^2\,+\,\frac{M_B}{2}B^2\,+\,
M_{{\cal{S}}B}{\cal{S}}B\,+\,\frac{M_T}{2}\tau_0^2\,.$$ The
corresponding terms for charged fermion masses are
$M_T\,\tau_+\tau_-\,-v_uY_i^T\,e_i\tau_+\,.$ The full
neutrino mass matrix, in an
$(\nu_i,\,{\cal{S}},\,{{B}},\,\tau_0)$-basis, is
\begin{equation}
\,{\cal{M}}_N\,=\,\left(\begin{array}{cc}
0\,&\,{\cal{M}}_D\\
\,&\,\\
{\cal{M}}_D^{\bot}\,&\,{\cal{M}}_R
\end{array}\right)\,,
\end{equation}
where
$${\cal{M}}_D\,=\,v_u\,\left(\begin{array}{ccc}
Y_1^{\cal{S}}\,&\,Y_1^{B}\,&\,-\frac{1}{\sqrt{2}}Y_1^T\,\\
\,&\,&\,\\
Y_2^{\cal{S}}\,&\,Y_2^{B}\,&\,-\frac{1}{\sqrt{2}}Y_2^T\,\\
\,&\,&\,\\
Y_3^{\cal{S}}\,&\,Y_3^{B}\,&\,-\frac{1}{\sqrt{2}}Y_3^T\,
\end{array}\right),\,\,\,\,\,\,\,\,\,
{\cal{M}}_R\,=\,\left(\begin{array}{ccc}
M_{\cal{S}}\,&\,M_{{\cal{S}}B}\,&\,0\\
\,&\,&\,\\
M_{{\cal{S}}B}\,&\,M_B\,&\,0\\
\,&\,&\,\\
0\,&\,0\,&\,M_T
\end{array}\right)\,.$$
Note that $Y_i^B\,=\,\frac{3}{\sqrt{30}}Y_i^T$.
The constraints on $\mu'$ and $f$ imply that
$M_B\,\approx\,M_G$, while $M_{\cal{S}}$ and
$M_{{\cal{S}}B}$ remain undetermined.

The light neutrino mass
matrix will be
\begin{equation}{\cal{M}}_{\nu}\,\approx\,-{\cal{M}}_D\,{\cal{M}}_R^{-1}\,{\cal{M}}_D^{\bot}\,.\end{equation}
The inverse right-handed neutrino Majorana mass is
\begin{equation}{\cal{M}}_R^{-1}\,=\,\frac{1}{\Delta}\left(\begin{array}{ccc}
-M_B\,&\,M_{{\cal{S}}B}\,&\,0\\
\,&\,&\,\\
M_{{\cal{S}}B}\,&\,-M_{{\cal{S}}}\,&\,0\\
\,&\,&\,\\
0\,&\,0\,&\,\frac{\Delta}{M_T}
\end{array}\right)\,,\end{equation}
with $\Delta\,=\,M_{{\cal{S}}B}^2\,-M_{\cal{S}}M_B$.

The determinant of ${\cal{M}}_D$ vanishes due to the $SU(5)$
relation $Y_i^T\,=\,\frac{\sqrt{30}}{3}Y_i^B$. This propagates to
${\cal{M}}_{\nu}$ resulting in one massless left-handed neutrino.
 Such a feature
is shared by a wider class of models in which two right-handed
neutrinos or more belong to the same GUT representation.

The resulting light neutrino mass
matrix can be put in the form
\begin{equation}\left({\cal{M}}_{\nu}\right)_{ij}\,=\,\frac{v_u^2}{\Delta}\left(\,A\,Y_iY_j\,+\,B\,\left(Y_iY_j'\,+\,Y_i'Y_j\right)\,+\,C\,Y_i'Y_j'\right)\,.{\label{NEU-1}}\end{equation}
where
\begin{equation}A\,=\,M_B,\,\,B\,=\,-\frac{3}{\sqrt{30}}M_{{\cal{S}}B},\,\,\,\,C\,=\,\frac{3}{10}M_{\cal{S}}\,-\frac{\Delta}{2M_T}\,.{\label{C}}\end{equation}
We have simplified the notation by denoting $Y_i^{\cal{S}}=Y_i$ and $Y_i^T=Y_i'$.

By going to the orthogonal basis in flavor space
\begin{equation}\hat{X}^{(1)}\,=\,\frac{{\vec{Y}^{\,\prime}\times \vec{Y}\,\,}}{|\vec{Y}^{\,\prime}\times\vec{Y}|},\,\,\hat{X}^{(2)}\,=\,\frac{\vec{Y}}{\sqrt{Y^2}},\,\,\,
\,\hat{X}^{(3)}\,=\,\hat{X}^{(1)}\times\hat{X}^{(2)},\,{\label{XXX}}\end{equation}
where $\hat{X}^{(1)}$ is the massless eigenvector, we can set the
neutrino matrix in the form
\begin{equation}{\cal{M}}_{\nu}\,=\,\left(\begin{array}{ccc}
0\,&\,0\,&\,0\\
\,&\,&\,\\
0\,&\,M_{22}\,&\,M_{23}\\
\,&\,&\,\\
0\,&\,M_{23}\,&\,M_{33}
\end{array}\right)\,,{\label{NEU-2}}\end{equation}
with
$$M_{22}\,=\,\frac{v_u^2}{\Delta}\left(\,M_B\,Y^2\,-\frac{6}{\sqrt{30}}M_{{\cal{S}}B}(Y\cdot Y')\,+\,\frac{3}{10}M_{{\cal{S}}}\frac{(Y\cdot Y')^2}{Y^2}\,\right)\,-\frac{v_u^2}{2M_T}\frac{(Y\cdot Y')^2}{Y^2}$$
$$M_{23}\,=\,\sqrt{Y^2{Y'}^2\,-(Y\cdot Y')^2}\left\{\,-\frac{v_u^2}{\Delta}\left(\,-\frac{3}{\sqrt{30}}M_{{\cal{S}}B}\,+\,\frac{3}{10}M_{{\cal{S}}}\frac{(Y\cdot Y')}{Y^2}\,\right)\,+\,\frac{v_u^2}{2M_T}\frac{(Y\cdot Y')}{Y^2}\right\}$$
\begin{equation}\,\,\,M_{33}\,=\,\frac{1}{Y^2}\left(\,{Y'}^2\,Y^2\,-(Y\cdot Y')^2\,\right)\left\{\,-\frac{v_u^2}{2M_T}\,+\,\frac{3v_u^2}{10}\frac{M_{\cal{S}}}{\Delta}\,\right\}\,.\end{equation}
Before we extract the light neutrino eigenvalues from this matrix, we must consider the scales involved in these expressions. For the mass scale $M_B$ we have already made the choice $M_B=M_G$. The other two scales $M_{\cal{S}},\,M_{{\cal{S}}B }$, associated with the singlet ${\cal{S}}$, are not constrained.

{\textit{\textbf{1st Approach:}}} We shall assume that these two
scales are also of the order of $M_G$. Thus, the dominant entry in
the neutrino matrix elements $M_{ab}$ will be the term
$-\frac{v_u^2}{M_T}$ contained in $C$ of ({\ref{C}}), while the
rest of the contributions will all be of the order of
$\frac{v_u^2}{M_G}$, which is three orders of magnitude smaller.
We may write\footnote{We have set
$$\hat{M}_{22}\,=\,\frac{1}{\sqrt{{|\Delta|}}}\left(\,M_BY^2-\frac{6M_{{\cal{S}}B}}{\sqrt{30}}(Y\cdot Y')\,+\,\frac{3M_{{\cal{S}}}}{10}\frac{(Y\cdot Y')^2}{Y^2}\right)$$
$$\hat{M}_{23}\,=\,\frac{1}{\sqrt{{|\Delta|}}}\sqrt{Y^2{Y'}^2-(Y\cdot Y')^2}\,\left(\,\frac{3M_{{\cal{S}}B}}{\sqrt{30}}\,-\frac{3M_{\cal{S}}}{10}\frac{(Y\cdot Y')}{Y^2}\right),\,\,\hat{M}_{33}\,=\,\frac{3M_{\cal{S}}}{10\sqrt{{|\Delta|}}}\left(\,{Y'}^2\,-\frac{(Y\cdot Y')^2}{Y^2}\right)\,.$$}
$$M_{22}\,=\,\frac{v_u^2}{\sqrt{{|\Delta|}}}\hat{M}_{22}\,-\frac{v_u^2}{2M_T}\frac{(Y\cdot Y')^2}{Y^2}$$
$$M_{23}\,=\,\frac{v_u^2}{\sqrt{{|\Delta|}}}\hat{M}_{23}\,+\,\frac{v_u^2}{2M_T}\frac{(Y\cdot Y')}{Y^2}\sqrt{Y^2{Y'}^2\,-(Y\cdot Y')^2}$$
\begin{equation}M_{33}\,=\,\frac{v_u^2}{\sqrt{{|\Delta|}}}\hat{M}_{33}\,-\frac{v_u^2}{2M_T}\,\frac{1}{Y^2}\,\left(\,{Y'}^2\,Y^2\,-(Y\cdot Y')^2\,\right)\,.\end{equation}
The resulting light neutrino mass eigenvalues are
$$m_{\nu}^{(3)}\,\approx\,-\frac{v_u^2}{2{M_T}}{Y'}^2,\,\,\,\,\,\,\,\,m_{\nu}^{(2)}\,\approx\,\frac{v_u^2}{\sqrt{{|\Delta|}}}\left\{\,\hat{M}_{22}\left(\,1\,-\frac{(Y\cdot Y')^2}{Y^2{Y'}^2}\,\right)\,+\,\hat{M}_{33}\frac{(Y\cdot Y')^2}{Y^2{Y'}^2}\,+\,\right.$$
\begin{equation}\left.\,2\hat{M}_{23}\frac{(Y\cdot Y')}{Y^2{Y'}^2}\sqrt{Y^2{Y'}^2-(Y\cdot Y')^2}\,\right\}\,.\end{equation}
As it stands, for ${|Y|\,\sim\,|Y'|}$, the mass hierarchy is
$m_{\nu}^{(2)}/m_{\nu}^{(3)}\,\sim\,\frac{v^2}{M_G}O(Y^2)/\frac{v^2}{M_T}O(Y^2)\,\sim\,M_T/M_G\,\sim\,\epsilon\,,$
 which is too strong a hierarchy to satisfy the data, without any other adjustment of parameters. On the other hand, if the overall scale of the determinant $\sqrt{{|\Delta|}}\,=\,\sqrt{|M_{{\cal{S}}B}^2-M_{{\cal{S}}}M_B|}$ is set to be $\sqrt{{|\Delta|}}\,\sim\,\lambda\,M_G$, with $\lambda\,\sim\,{\mathcal{O}(10^{-1})}$, the relation $v^2/M_T>>v^2\hat{M}_{ab}/\sqrt{|\Delta|}$ still holds and, thus, we obtain
 \begin{equation}m_{\nu}^{(2)}\,\sim\,\frac{v_u^2}{\lambda^2M_G}\,O(Y^2),\,\,\,\,\,\,\,m_{\nu}^{(3)}\,\sim\,\frac{v_u^2}{M_T}\,O(Y^2)\,.\end{equation}
 This can give the correct overall scale of the neutrino masses and a suitable hierarchy
 \begin{equation}\frac{m_{\nu}^{(2)}}{m_{\nu}^{(3)}}\,\sim\,\frac{\epsilon}{\lambda^2}\,.\end{equation}

{\textit{\textbf{2nd Approach:}}} An alternative assumption is to
assume that the scales associated with the singlet
${\cal{S}}$ are of the the same intermediate order as $M_T$,
namely
\begin{equation}M_{\cal{S}}\,\sim\,M_{{\cal{S}}B }\,\sim\,M_T\,\end{equation}
and, thus, $\Delta\approx-M_{\Su}M_B$. Despite naturalness objections, this assumption is technically
feasible. In this case, we have to leading order
\begin{equation}
m_{\nu}^{(2,\,3)}\,\approx\,-\frac{v_u^2}{4M_T}\left\{\,(Y')^2+\lambda'
Y^2\,\pm\sqrt{\cal{R}}\,\right\}\,,\end{equation} where
\begin{equation}
{\cal{R}}\,\equiv\,\left(\,\lambda'\,Y^2\,-{Y'}^2\right)^2\,+\,4\lambda'\,(Y\cdot
Y')^2\,
\end{equation}
and $\lambda '\equiv 2M_T/M_{\cal{S}}$, a number of $O(1)$ by
assumption. Note that a hierarchy can also arise in this approach in the case $Y^2\,>>\,{Y'}^2$, namely
\begin{equation}
\frac{m_{\nu}^{(2)}}{m_{\nu}^{(3)}}\,\approx\,\frac{\left(Y^2{Y'}^2\,-(Y\cdot
Y')^2\,\right)}{\lambda'
Y^4}\,=\,{\frac{Y'^2\sin^2{\alpha}}{\lambda'
Y^2}}\,.\end{equation} We have denoted by $\alpha$ the angle
${\cos^{-1}(\hat{Y}\cdot \hat{Y'}).}$ Similar results can also be
obtained  for $Y'^{2}>>Y^{2}$ but with
\begin{equation}
\frac{m_{\nu}^{(2)}}{m_{\nu}^{(3)}}\,\approx\,\,{\frac{\lambda'
Y^2\sin^2{\alpha}}{Y'^2}}\,.\label{2b}
\end{equation}

In this approach there is also another possibility for the
existence of a mass-hierarchy, namely, the possibility of almost
parallel couplings in generation space
($\alpha\,{\approx\,0}$)
$$Y\cdot Y'\,=\,\sqrt{Y^2}\,\sqrt{{Y'}^2}\,\cos\alpha\,\approx\,\sqrt{Y^2}\,\sqrt{{Y'}^2}\,\left(\,1\,-\frac{\alpha^2}{2}\,\right)\,.$$
In this case, keeping $Y\,\sim\,Y'$, we obtain
\begin{equation}
\frac{m_{\nu}^{(2)}}{m_{\nu}^{(3)}}\,\approx\,\frac{{\lambda'}Y^2{Y'}^2\,\alpha^2}{({Y'}^2+\lambda' Y^2)^2}\,.
\end{equation}
Finally, in this approach, there is a third possibility for a
hierarchy if we assume that there is a small hierarchy in the
scales $M_S:M_T$ corresponding to {$\lambda'\,\sim\,0.1$. In
this case we get the same expression for the mass ratio as in
(\ref{2b}) but with the desired hierarchy now originating from
$\lambda'$ instead of $Y^2/Y'^2$.}

The above conclusions rely only on the renormalizable part of the superpotential. There are however some contributions to neutrino masses from various lowest order non-renormalizable terms in ({\ref{NRW}}). These are:\\
{\textit{Left-handed neutrino Majorana masses}} from the term
\begin{equation}
{\cal{H}}^c{\cal{Q}}{\cal{Q}}{\cal{H}}^c\,\sim\,\lambda_{ij}\frac{v_u^2}{M_P}\,\nu_i\,\nu_j\,.
\end{equation}
These masses are tiny ($\,10^{-5}\,eV$ or less, depending on the couplings involved) but they remove the massless state arising from the previous analysis giving a lower bound for light neutrino masses.\\
{\textit{Right-handed neutrino Majorana masses}} from the terms
\begin{equation}
{\cal{T}}^2\Sigma^2\,+\,\Sigma^2{\cal{T}}{\cal{S}}\,+\,{\cal{S}}^2\Sigma^2\,\sim\,\lambda_{ij}'\,\frac{V^2}{M_P}\,N_j^cN_j^c\,.
\end{equation}
These terms could very well be of the same order of magnitude as the intermediate scale $M_T$ or even larger but become subdominant  for relatively  small couplings, meaning $\lambda'< 10^{-2}$.
 In addition to these terms, negligible right-handed Majorana mass contributions $O(v^2/M_P)$ arise from the
 terms ${\cal{H}}\left(\,{\cal{T}}^2,\,{\cal{T}}{\cal{S}},\,{\cal{S}}^2\right){\cal{H}}^c\,.$\\
{\textit{Dirac neutrino masses}} from the terms
\begin{equation}
{\cal{Q}}{\cal{T}}\Sigma{\cal{H}}^c\,+\,{\cal{Q}}\Sigma{\cal{H}}^c{\cal{S}}\,\sim\,\lambda_{ij}'\,\frac{v_uV}{M_P}\,\nu_i\,N_j^c\,.
\end{equation}
{These contributions, suppressed by the factor $V/M_P$ in comparison with renormalizable contributions,
 can remove massless states that arise due to the symmetries encountered in the renormalizable part of
 the Dirac neutrino mass matrix ${\cal{M}}_D$. To be specific, the operator ${\cal{Q}}{\cal{T}}\Sigma{\cal{H}}^c$
 representing the  invariants ${\cal{Q}}_i{\cal{H}}^cTr({\cal{T}}\Sigma),\,{\cal{Q}}_i{\cal{T}}\Sigma{\cal{H}}^c,\,{\cal{Q}}_i\Sigma{\cal{T}}{\cal{H}}^c$
 contributes to the superpotential as
\begin{equation}\lambda_{1i}''\frac{v_uV}{M_P}\nu_iB\,+\,(\lambda_{2i}''+\,\lambda_{3i}'')\,\frac{v_uV}{M_P}\left(\,\frac{3}{10}\nu_iB\,-\sqrt{\frac{3}{20}}\,\nu_i\,{\cal{\tau}}_0\,\right)\,.
\end{equation}
The presence of these terms modifies the structure of
${\cal{M}}_D$ and removes the massless state. The resulting from
the seesaw mechanism light neutrino mass will be suppressed at
least by a factor of $(\lambda''V/M_P)^2\,<\,10^{-2}$ compared to
the lightest massive neutrino.}

\section{Neutrino Mixing}
The charged lepton and neutrino mass terms
{$M_{(\ell)}\,\ell\,\ell^c\,+\frac{1}{2}M_{(\nu)}\,\nu\,\nu$}
can be diagonalized in terms of three unitary matrices
${\bf{U}}_{(\ell)}$, ${\bf{V}}_{(\ell^c)}$ and ${\bf{U}}_{(\nu)}$.
These matrices rotate the above {\textit{gauge eigenstates}} into
{\textit{mass eigenstates}}. If we express the neutrino charge
current $J_{\mu}\propto \ell^{\dagger}\sigma_{\mu}\nu\,$ in terms
of mass eigenstates, a combination of two of these matrices will
appear ${\ell^{\dagger}}'\,\sigma_{\mu}\,{\cal{U}}_{PMNS}\,\nu'$,
known as the {\textit{Pontecorvo-Maki-Nakagawa-Sakata}}\cite{PMNS}
matrix
\begin{equation}
{\cal{U}}_{PMNS}\,\equiv\,{\bf{U}}_{(\ell)}^{\dagger}\,{\bf{U}}_{(\nu)}\,.
\end{equation}
In what follows we shall concentrate on ${\bf{U}}_{(\nu)}$ and put aside the charged lepton mixing matrix, for which, in any case very little is known.

The overall neutrino mixing matrix
\begin{equation}
{\bf{U}}_{(\nu)}\,=\,{\bf{U}}_1\,{\bf{U}}_2,\,\,\,\,\,\,\,\,\,\,\,\left(\,\left({\bf{U}}_1\right)_{ij}\,=\,\hat{X}_j^{(i)}\,\right)
 \end{equation}
 is composed of the unitary matrix  ${\bf{U}}_1$ that rotates the neutrino mass matrix ({\ref{NEU-1}) into ({\ref{NEU-2}}) and a unitary matrix
\begin{equation}
{\bf{U}}_2\,=\,\left(\begin{array}{ccc}
1\,&\,0\,&\,0\\
0\,&\,\cos\beta\,&\,-\sin\beta\\
0\,&\,\sin\beta\,&\,\cos\beta
\end{array}\right)
\end{equation}
that diagonalizes ({\ref{NEU-2}}). The rotation angle $\beta$ is related to the matrix entries through
\begin{equation}\beta\,\equiv\,\frac{1}{2}\cot^{-1}\left(\frac{M_{22}-M_{33}}{2M_{23}}\right)\,.\end{equation}
 Note that the mass
eigenvalues are just
$$m_{\nu}^{(2,3)}\,=\,\frac{1}{2}\left(M_{22}+M_{33}\,\pm\sqrt{(M_{22}-M_{33})^2+4M_{23}^2}\right)\,.$$
The overall diagonalizing matrix is
\begin{equation}
{\bf{U}}_{\nu}\,=\,{\bf{U}}_1\,{\bf{U}}_2\,=\,\left(\begin{array}{ccc}
\hat{X}_1^1\,&\,\cos\beta\,\hat{X}_2^1+\sin\beta\,\hat{X}_3^1\,&\,-\sin\beta\,\hat{X}_2^1\,+\,\cos\beta\,\hat{X}_3^1\\
\,&\,&\,\\
\hat{X}_1^2\,&\,\cos\beta\,\hat{X}_2^2+\sin\beta\,\hat{X}_3^2\,&\,-\sin\beta\,\hat{X}_2^2\,+\,\cos\beta\,\hat{X}_3^2\\
\,&\,&\,\\
\hat{X}_1^3\,&\,\cos\beta\,\hat{X}_2^3+\sin\beta\,\hat{X}_3^3\,&\,-\sin\beta\,\hat{X}_2^3\,+\,\cos\beta\,\hat{X}_3^3
\end{array}\right)
\end{equation}
In order to obtain the corresponding relations between the $\hat{X}^{(a)}$ and the original Yukawa couplings $Y_i$ and $Y_i'$, we note that, as a result of the definitions ({\ref{XXX}}), we may write
$$\vec{Y}^{\,\prime}\,=\,Y'\left(\,\cos\alpha\,{\hat{X}_2}\,-\sin\alpha\,{\hat{X}_3}\,\right)\,,$$
where $\alpha\,\equiv\,\cos^{-1}\left(\hat{Y}\cdot\hat{Y}^{\,\,\prime}\right)\,\,.$ Substituting, we obtain
\begin{equation} {\bf{U}}_{\nu}=(\sin\alpha)^{-1}\left(\begin{array}{ccc}
\hat{Y}_3\hat{Y}_2'-\hat{Y}_2\hat{Y}_3'\,&\,\sin(\alpha{+}\beta)\hat{Y}_1-\sin\beta\hat{Y}_1'\,&\,\cos(\alpha{+}\beta)\hat{Y}_1-\cos\beta\hat{Y}_1'\\
\,&\,&\,\\
\hat{Y}_3'\hat{Y}_1-\hat{Y}_1'\hat{Y}_3\,&\,\sin(\alpha{+}\beta)\hat{Y}_2-\sin\beta\hat{Y}_2'\,&\,\cos(\alpha{+}\beta)\hat{Y}_2-\cos\beta\hat{Y}_2'\\
\,&\,&\,\\
\hat{Y}_2\hat{Y}_1'-\hat{Y}_1\hat{Y}_2'\,&\,\sin(\alpha{+}\beta)\hat{Y}_3-\sin\beta\hat{Y}_3'\,&\,\cos(\alpha{+}\beta)\hat{Y}_3-\cos\beta\hat{Y}_3'
\end{array}\right)\,.{\label{MIXMAT}}\end{equation}
Equating this matrix with the standard parametrization we
obtain the relations between the standard mixing angles
$\theta_{23},\,\theta_{12},\,\theta_{13}$ and the above
parameters. It is clear that, as long as we have not imposed any
additional constraints on the Yukawa coupling directions in family
space, we have no predictive restrictions on the mixing angles. In
the particular case that we are close to {\textit{bimaximal
mixing}}
$$\theta_{23}\,\approx\,\frac{\pi}{4}\,+\,\epsilon_{23},\,\,\,\,\,\,\,\,\,\,\,\,\theta_{12}\,\approx\,\frac{\pi}{4}\,+\,\epsilon_{12},\,\,\,\,\,\,\,\,\,\,\theta_{13}\,\approx\,\epsilon_{13}\,,$$
 from the standard parametrization we obtain
\begin{equation}{\bf{U}}_{(\nu)}\,\approx\,\left(\begin{array}{ccc}
\frac{1}{\sqrt{2}}-\frac{\epsilon_{12}}{\sqrt{2}}\,&\,\frac{1}{\sqrt{2}}\,+\,\frac{\epsilon_{12}}{\sqrt{2}}\,&\,\epsilon_{13}\\
\,&\,&\,\\
-\frac{1}{2}\,-\frac{\epsilon_{12}}{2}\,+\,\frac{\epsilon_{23}}{2}\,-\frac{\epsilon_{13}}{2}\,&\,\frac{1}{2}\,-\frac{\epsilon_{12}}{2}\,-\frac{\epsilon_{23}}{2}\,-\frac{\epsilon_{13}}{2}\,&\,\frac{1}{\sqrt{2}}\,+\,\frac{\epsilon_{23}}{\sqrt{2}}\\
\,&\,&\,\\
\frac{1}{2}\,+\,\frac{\epsilon_{12}}{2}\,+\,\frac{\epsilon_{23}}{2}\,-\frac{\epsilon_{13}}{2}\,&\,-\frac{1}{2}\,+\,\frac{\epsilon_{12}}{2}\,-\frac{\epsilon_{23}}{2}\,-\frac{\epsilon_{13}}{2}\,&\,\frac{1}{\sqrt{2}}\,-\frac{\epsilon_{23}}{\sqrt{2}}
\end{array}\right)\,.\end{equation}
Equating this expression to ({\ref{MIXMAT}}), we obtain
\begin{equation}\hat{Y}\,=\,\left[\begin{array}{c}
\frac{\cos\beta}{\sqrt{2}}\\
\,\\
\frac{\cos\beta}{2}\,-\frac{\sin\beta}{\sqrt{2}}\\
\,\\
-\frac{\cos\beta}{2}\,-\frac{\sin\beta}{\sqrt{2}}
\end{array}\right]\,+\,\left[\begin{array}{c}
\epsilon_{12}\frac{\cos\beta}{\sqrt{2}}\,-\epsilon_{13}\sin\beta\\
\,\\
-\left(\epsilon_{12}+\epsilon_{23}+\epsilon_{13}\right)\frac{\cos\beta}{2}\,-\epsilon_{23}\frac{\sin\beta}{\sqrt{2}}\\
\,\\
-\left(-\epsilon_{12}+\epsilon_{23}+\epsilon_{13}\right)\frac{\cos\beta}{2}\,+\,\epsilon_{23}\frac{\sin\beta}{\sqrt{2}}
\end{array}\right]\end{equation}
and
\begin{equation}\hat{Y}'\,=\,\left[\begin{array}{c}
\frac{\cos(\alpha+\beta)}{\sqrt{2}}\\
\,\\
\frac{\cos(\alpha+\beta)}{2}\,-\frac{\sin(\alpha+\beta)}{\sqrt{2}}\\
\,\\
-\frac{\cos(\alpha+\beta)}{2}\,-\frac{\sin(\alpha+\beta)}{\sqrt{2}}
\end{array}\right]\,+\,\left[\begin{array}{c}
\epsilon_{12}\frac{\cos(\alpha+\beta)}{\sqrt{2}}\,-\epsilon_{13}\sin(\alpha+\beta)\\
\,\\
-\left(\epsilon_{12}+\epsilon_{23}+\epsilon_{13}\right)\frac{\cos(\alpha+\beta)}{2}\,-\epsilon_{23}\frac{\sin(\alpha+\beta)}{\sqrt{2}}\\
\,\\
-\left(-\epsilon_{12}+\epsilon_{23}+\epsilon_{13}\right)\frac{\cos(\alpha+\beta)}{2}\,+\,\epsilon_{23}\frac{\sin(\alpha+\beta)}{\sqrt{2}}
\end{array}\right]\,.\end{equation}
Closing this chapter we note that the
range of values for variables $\alpha,\beta,|Y|,|Y'|$, which determine the
Yukawa couplings, depends on the
mass hierarchy approach followed. Among the different options, the small angle scenario of the \textit{2nd approach }
exhibits the most restrictive structure with $\beta\sim \alpha$,
while by assumption $|Y|\sim\,|Y'|$.

\section{Conclusions}

In the present article we studied a realization of the
see-saw mechanism  in the framework of an extended
renormalizable version of the supersymmetric $SU(5)$
model. The right-handed neutrino fields
were introduced as members of chiral ${\bf{24}}+{\bf{1}}$
superfields. In particular, two ${\bf{24}}$ superfields were
introduced, out of which, due to different discrete symmetry
charges, only one couples to matter and its neutral singlet and
isotriplet components are identified as two of the right-handed
neutrinos. Our basic assumption is that right-handed neutrinos
survive below the grand unification scale having an intermediate
mass in the neighborhood of $10^{13}-10^{14}\,GeV$, a scale
suitable to generate, through the see-saw mechanism, a light
neutrino mass of the observed mass value of $O(0.1\,GeV)$. The
assumption of an isotriplet of an intermediate mass scale is
supported by renormalization group analysis incorporating proton
stability constraints. In addition, the model requires a color octet of neighboring mass, which, however, does not couple to ordinary matter. The right-handed neutrino mass matrix, then, depends on the constrained isotriplet scale $M_T$ as well as the free, from renormalization group,  scales  $M_B,M_\mathcal{S} , M_{\mathcal{S}B}$ 
associated with the SM singlets of $\mathbf{\underline{1},\underline{24}}$. If these scales are of $O(M_G)$, an extra fine
tuning is required in order to obtain a light neutrino mass
hierarchy in agreement with data
({\textit{1st approach}}). The alternative assumption according to
which the scales $M_{\Su},\,M_{SB}$ are of
$O(M_T)$ is also possible ({\textit{2nd approach}}). In this
approach a phenomenologically acceptable neutrino mass hierarchy
is possible as a result of the Yukawa hierarchy $Y'\,<<\,Y$
or $Y'\,>>\,Y$, where $Y$ and $Y'$ are the overall scales of
the neutrino couplings
$<H_u>\nu\,\left(\,Y\,{\bf{1}}\,+\,Y'\,{\bf{24}}\,\right)$. A
second possibility of a hierarchy within this approach arises also
when the angle between the Yukawa coupling vectors in family space
$Y_i$ and $Y_i'$ is small. Nevertheless, the limiting case of
aligned Yukawas is excluded, since it corresponds to two massless
neutrinos. Alternatively, the required neutrino mass hierarchy can
also arise as a result of a slight hierarchy of the scales
$M_S:M_T$. However, in all these approaches, one
very light neutrino is always present as a result of the structure
of the neutrino mass matrix. Finally, we also find that
a hierarchical mixing angle structure
$\theta_{23}\,\sim\,\theta_{12}\,>>\,\theta_{13}$ can be easily
accommodated within the free parameter structure of the model.

{\textbf{Acknowledgements}}

The research presented in this article is co-funded by the European Union - European Social Fund (ESF) $\&$ National Sources, in the framework of the Programme {\textit{"HRAKLEITOS II"}} of the {\textit{"Operational Programme for Education and Lifelong Learning"}} of the {\textit{Hellenic Ministry of Education, Lifelong Learning and Religious Affairs}}. Both authors acknowledge the hospitality of the CERN Theory Group.

\end{document}